\documentclass[aps,prb,twocolumn,superscriptaddress]{revtex4}
\usepackage{graphicx}
\usepackage{amssymb}
\usepackage{amsmath}
\usepackage{graphicx}
\usepackage{epsfig}
\usepackage{color}
\usepackage{natbib}

\begin{document}

\title{Self-doping and the Mott-Kondo scenario for infinite-layer nickelate
superconductors}
\author{Yi-feng Yang}
\email[]{yifeng@iphy.ac.cn}
\affiliation{Beijing National Laboratory for Condensed Matter Physics and Institute of
Physics, Chinese Academy of Sciences, Beijing 100190, China}
\affiliation{School of Physical Sciences, University of Chinese Academy of Sciences,
Beijing 100190, China}
\affiliation{Songshan Lake Materials Laboratory, Dongguan, Guangdong 523808, China}
\author{Guang-Ming Zhang}
\email[]{gmzhang@tsinghua.edu.cn}
\affiliation{State Key Laboratory of Low-Dimensional Quantum Physics and Department of
Physics, Tsinghua University, Beijing 100084, China}
\affiliation{Frontier Science Center for Quantum Information, Beijing 100084, China}
\date{\today }

\begin{abstract}
We give a brief review of the Mott-Kondo scenario and its consequence in the
recently-discovered infinite-layer nickelate superconductors. We argue that
the parent state is a self-doped Mott insulator and propose an effective $t$-%
$J$-$K$ model to account for its low-energy properties. At small doping, the
model describes a low carrier density Kondo system with incoherent Kondo
scattering at finite temperatures, in good agreement with experimental
observation of the logarithmic temperature dependence of electric
resistivity. Upon increasing Sr doping, the model predicts a breakdown of
the Kondo effect, which provides a potential explanation of the non-Fermi
liquid behavior of the electric resistivity with a power law scaling over a
wide range of the temperature. Unconventional superconductivity is shown to
undergo a transition from nodeless $(d+is)$-wave to nodal $d$-wave near the
critical doping due to competition of the Kondo and Heisenberg superexchange
interactions. The presence of different pairing symmetry may be supported by
recent tunneling measurements.
\end{abstract}

\maketitle

\section{Introduction}

Recent discovery of superconductivity (SC) in infinite-layer Sr-doped NdNiO$%
_{2}$ films \cite{Li2019} and subsequently in hole doped LaNiO$_2$ and PrNiO$%
_2$ films \cite{Osada2020a,Osada2020b,Zeng2021,Osada2021} has stimulated
intensive interest in condensed matter community. Despite of many
theoretical and experimental efforts, there are still debates on its
electronic structures and pairing mechanism \citep%
{zhang-yang-zhang,Wang2020,Botana2020,Jiang2020,Sakakibara2020,Hepting2020,Normura2019,Gao2021}%
. The study of possible Ni-based superconductivity was initially stimulated
by cuprates, whose high $T_c$ mechanism remains one of the most challenging
topics in past three decades \cite{Muller,Anderson,AtoZ,LeeNagaosaWen}. Many
attempts have been devoted to exploring new families of high $T_c$
superconductors. Nickelate superconductors are but one latest example of
these efforts.

In undoped cuprates, Cu$^{2+}$ ions contain 9 electrons with partially
occupied $3d_{x^{2}-y^{2}}$ orbitals. The oxygen $2p$ orbitals are higher in
energy than the Cu $3d_{x^{2}-y^{2}}$ lower Hubbard band. Thus, cuprates
belong to the so-called charge-transfer insulator. A superexchange
interaction between localized Cu $3d_{x^{2}-y^{2}}$ spins is mediated by
oxygen ions and causes an antiferromagnetic (AF) ground state. Upon chemical
doping, holes may be introduced on the oxygen sites in the CuO$_{2}$ planes
\cite{Anderson,AtoZ,LeeNagaosaWen} and combine with the $3d_{x^{2}-y^{2}}$
spins to form the Zhang-Rice singlets \cite{ZhangRice}, destroying the
long-range AF order rapidly. In theory, these led to an effective $t$-$J$ model, describing the holes moving on the
antiferromagnetic square lattice. High temperature SC with robust $d$-wave
pairing has been predicted and established over a wide doping range \cite%
{Shen,Harlingen,Tsuei}. Extending such \textquotedblleft cuprate-Mott"
conditions in other oxides has led to extensive efforts on nickel oxides
\cite%
{Anisimov1999,Hayward1999,Lee2004,Botana2017,Chaloupka2008,Hansmann2009,Middey2016,Boris2011,Benckiser2011,Disa2015,Zhang2017}%
. Nickelate superconductors have a similar layered crystal structure with Ni$%
^{1+}$ possessing the same $3d^{9}$ configuration as Cu$^{2+}$. As a result,
theories based on Mott scenario have naturally been developed to account for
nicklate superconductors.

\begin{figure}[t]
\centerline{{\includegraphics[width=.45\textwidth]{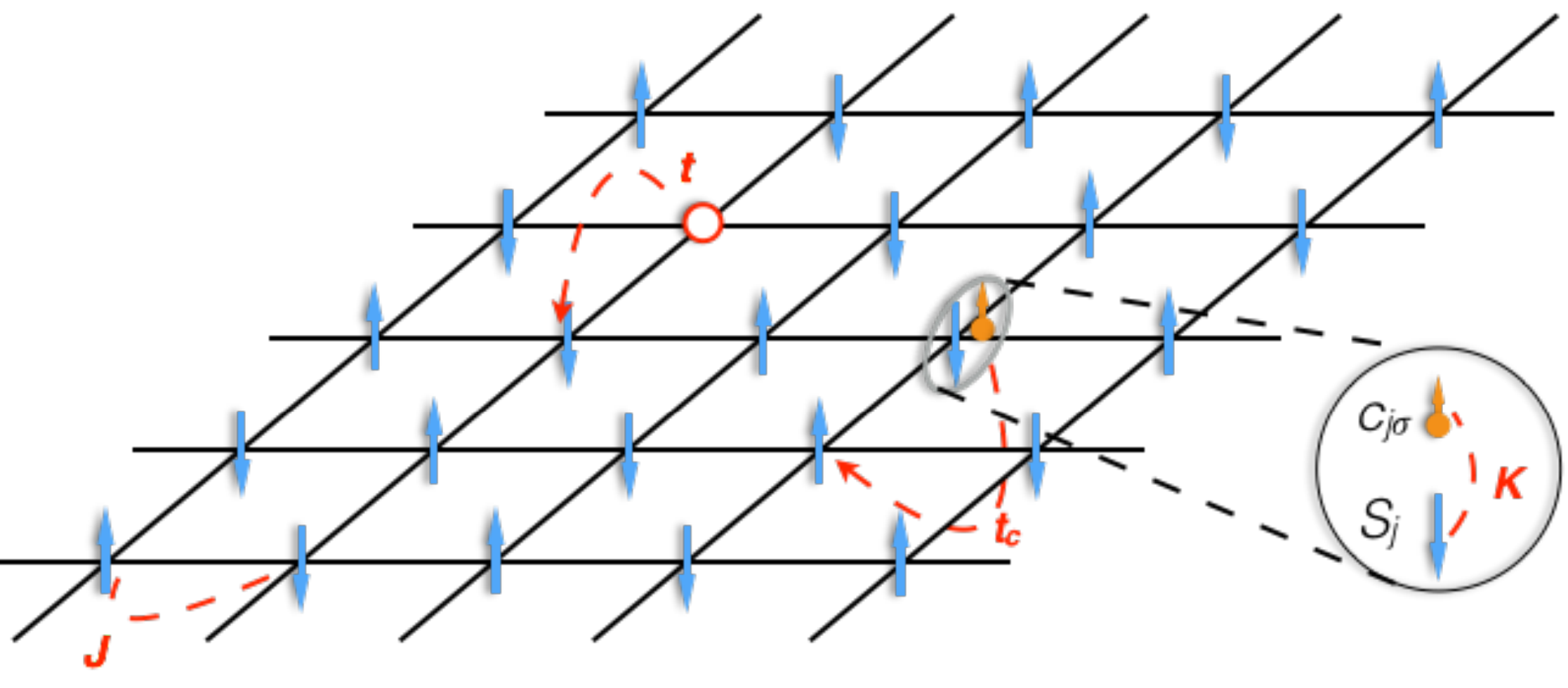}}}
\caption{Illustration of the effective model on a two-dimensional square
lattice of NiO$_{2}$ plane of NdNiO$_{2}$. Blue arrow represents Ni-spin,
which interacts with its neighboring spin antiferromagetically by coupling $%
J $. Orange arrow denotes Nd-$5d$ electron, which couples to Ni-spin by the
Kondo coupling $K$, to form a Kondo singlet (doublon). Red circle represents
Ni-$3d^{8}$ configuration, or a holon. $t_c$ and $t$ denote the hopping of doublon and holon, respectively. Not shown is the holon-doublon
anhilation into a Ni-spin. Figure adapted from Ref. \protect\cite%
{zhang-yang-zhang}. Copyright 2020 by the American Physical Society.}
\label{figLattice}
\end{figure}

However, there are clear evidences since the beginning suggesting that these
two systems are different. Instead of a Mott insulator with AF long-range
order like in cuprates, NdNiO$_{2}$ displays metallic behavior at high
temperatures with a resistivity upturn below about $70$ K, showing no sign
of any magnetic long-range order in the whole measured temperature range
\cite{Hayward2003}. Similar results have previously been found in LaNiO$_{2}$
(Ref.\cite{Ikeda2016}). First-principles calculations have also revealed
some subtle differences in their band structures. The O-$2p$ orbitals are
located at a deeper energy compared to that of cuprates. Nd-5$d$ bands are
found to hybridize with Ni-$3d$ bands and produce small electron pockets in
the Brillouin zone. As a consequence, holes are doped directly into Ni-3$d$
orbitals rather than O-$2p$ orbitals. Nickelates should thus be modelled as
a self-doped Mott insulator, which implies a multi-band system with two
types of charge carriers, the itinerant Nd-$5d$ conduction electrons and the
Ni-3$d_{x^{2}-y^{2}}$ holes, on a background lattice of Ni-3$d_{x^{2}-y^{2}}$
magnetic moments \cite{zhang-yang-zhang}. Joint analysis of the resistivity
upturn and Hall coefficient at low temperatures suggests possible presence
of incoherent Kondo scattering between low-density conduction electrons and
localized Ni spins. A physical picture is illustrated in Fig.~\ref{figLattice}
on the square lattice. Then the basis for the Mott-Kondo scenario of nickelate 
superconductors has been established, leading to the proposal of an extended $t$-$J$-$K$
model for a microscopic description of their low-energy properties \cite%
{zhang-yang-zhang,Wang2020}.

Interestingly, the Kondo hybridization does not appear significant at first
glance in band structure calculations \cite{Lee2004}. It was later realized
that nickelates may host a special interstitial-$s$ orbital for conduction
electrons that have substantially stronger hybridization than previously
thought \cite{Hanghui2020}. Resonant inelastic X-ray scattering (RIXS)
measurements \cite{Hepting2020,Kourkoutis} confirmed the presence of
hybridization between Ni $3d_{x^{2}-y^{2}}$ and Nd $5d$ orbitals. At zero
temperature, the self-doping effect and the Kondo coupling produce
low-energy doublon (Kondo singlet) and holon excitations on the nickel
spin-1/2 background. Because of the larger charge transfer energy,
nickelates have a reduced superexchange interaction between Ni$^{1+}$ spins
by almost an order of magnitude than cuprates. Raman scattering measurements
confirmed this expectation and estimated $J\approx 25\,$meV in bulk NdNiO$%
_{2}$ \cite{Fu2019}. The Kondo coupling may therefore suppress the AF
long-range order and cause a phase transition to a paramagnetic metal \cite%
{zhang-yang-zhang}. The parent \ or underdoped compounds may therefore be
viewed as a Kondo semimetal (KS). For large hole doping, the Ni-3$d$
electrons become more itinerant and the Kondo effect breaks down, followed
by an abrupt change of the charge carriers. Indeed, a sign change of the
Hall coefficient has been reported in experiment \cite{Li2020,Zeng2020}.

The above differences have an immediate impact on candidate pairing
mechanism of the superconductivity. At critical doping, the $t$-$J$-$K$
model predicted possible SC transition from a gapped ($d+is$)-wave state to
a gapless $d$-wave pairing state due to the competition of Kondo and
superexchange interactions \cite{Wang2020}. Latest scanning tunneling
experiment (STM) also revealed two different gap structures of U and
V-shapes, supporting the possibility of above scenario. Thus, nickelates may
belong to a novel class of unconventional superconductors and one may
anticipate potentially more interesting properties bridging the cuprates and
heavy fermions.

In this paper, we briefly summarize the consequences of the Mott-Kondo
scenario based on the extended $t$-$J$-$K$ model for nickelate
superconductors \cite{zhang-yang-zhang,Wang2020}. We propose a global phase
diagram upon electron and hole doping and derive a low-energy effective
Hamiltonian with doublon and holon excitations in the low doping region. We
then employ the renormalized mean-field theory (RMFT) to study the
possibility of superconductivity and predict a phase transition of its
pairing symmetry. The latter is shown to originate from the breakdown of
Kondo hybridization, accompanied with non-Fermi liquid (NFL) behavior of the
resistivity $\rho\sim T^\alpha$ near critical doping.

\section{Theory}

\subsection{Model Hamiltonian}

To introduce the effective $t$-$J$-$K$ model for describing the low-energy
physics of nickelates, we start from a background lattice of Ni$^{1+}$ 3$%
d_{x^2-y^2}$ localized spins with a small number of self-doped holes and Nd-5%
$d$ conduction electrons \cite{zhang-yang-zhang}. Similar to cuprates, one
expects an AF superexchange interaction between Ni$^{1+}$ spins through the
O-2$p$ orbitals. This is different from heavy fermion systems, where the
exchange interaction between localized spins originates from the
Ruderman-Kittel-Kasuya-Yosida (RKKY) interaction mediated by conduction
electrons. The motion of holes on the spin lattice should be strongly
renormalized as in the usual $t$-$J$ model. There is an additional local
Kondo interaction between local spins and conduction electrons. The total
Hamiltonian therefore contains three terms:
\begin{equation}
H=H_{t}+H_{J}+H_{K},
\end{equation}
where the first term comes from the hopping of holes, the second term
describes the spin lattice, and the third term gives the Kondo interaction.

For simplicity, we consider a minimal model with Ni-$3d^{8}$ and Nd-$5d^{0}$
as the vacuum. As in cuprates, the localized $3d_{x^{2}-y^{2}}$ spins on the
NiO$_{2}$ plane can be described by a two-dimensional quantum Heisenberg
model with nearest neighbour AF superexchange interactions,
\begin{equation}
H_{J}=J\sum_{\langle ij\rangle }S_{i}\cdot S_{j},
\end{equation}
whose ground state is a Mott insulator with AF long-range orders. 

The self-doping effect is supported by first-principles band structure
calculations \cite{Lee-Pickett2004}, where the Nd $5d$ orbitals in NdNiO$%
_{2} $ are found to hybridize with the Ni $3d$ orbitals and give rise to
small electron pockets in the Brillouin zone. Thus, we have a small number
of Nd-5$d$ conduction electrons. This is actually supported by experiment.
At high temperatures, the Hall coefficient is dominated by conduction
electrons giving $R_H\approx -4\times 10^{-3}\,$cm$^{3}$\thinspace C$^{-1}$
for NdNiO$_{2}$ and $-3\times 10^{-3}\,$cm$^{3}$\thinspace C$^{-1}$ for LaNiO%
$_{2}$. By contrast, in typical heavy fermion metals such as Ce$M$In$_{5}$ ($%
M$= Co, Rh, Ir) , we have $R_{H}\approx -3.5\times 10^{-4}\,$cm$^{3}$%
\thinspace C$^{-1}$ at high temperatures \cite{Hundley2004}. The larger and
negative values of the Hall coefficient implies that there are only a few
percent of electron-like carriers per unit cell in NdNiO$_{2}$ and LaNiO$_{2}
$.

The hybridization between Ni 3$d_{x^2-y^2}$ spins and conduction electrons
gives the additional Kondo term:
\begin{equation}
H_{K}=-t_c\sum_{\langle ij\rangle ,\sigma }\left( c_{i\sigma }^{\dagger
}c_{j\sigma }+h.c.\right) +\frac{K}{2}\sum_{j\alpha ;\sigma \sigma ^{\prime
}}S_{j}^{\alpha }c_{j\sigma }^{\dagger }\tau _{\sigma \sigma ^{\prime
}}^{\alpha }c_{j\sigma ^{\prime }},
\end{equation}
where $\epsilon_{\mathbf{k}}$ is the dispersion of conduction electrons
projected on the square lattice of the Ni$^{1+}$ ions, and $\tau ^{\alpha }$
($\alpha =x,y,z$) are the spin-1/2 Pauli matrices. Only a single conduction
band is considered for simplicity. In reality, the pockets should be three
dimensional. For a low carrier density system, the average number of
conduction electrons is small, i.e. $n_c=N_{s}^{-1}\sum_{j\sigma }\langle
c_{j\sigma }^{\dagger }c_{j\sigma}\rangle \ll 1$.

The presence of magnetic impurities may be at first glance ascribed to the Nd $%
4f $ moments. However, the Nd$^{3+}$ ion contains three $f$ electrons
forming a localized spin-3/2 moment, which acts more like a classical spin
as in manganites and therefore disfavors spin-flip scattering as the quantum
spin-1/2 moment. Their energy level is also far away from the Fermi energy,
so it is reasonable to ignore the Nd $4f$ electrons.

For parent compounds, self-doping also introduces an equal number of Ni $%
3d_{x^2-y^2}$ holes on the spin lattice. The hopping of holes on the lattice of Ni $%
3d_{x^{2}-y^{2}}$ spins can be described as
interactions,
\begin{equation}
H_{t}=-\sum_{ij\sigma }\left( t_{ij}P_{G}d_{i\sigma }^{\dagger }d_{j\sigma
}P_{G}+h\text{.c.}\right),
\end{equation}%
where $d_{i\sigma }$ and $d_{i\sigma }^{\dag }$ are the annihilation and
creation operators of the Ni $3d_{x^{2}-y^{2}}$ electrons, respectively, $%
t_{ij}$ is the hopping integral between site $i$ and $j$, and $P_{G}$ is the
Gutzwiller operator to project out doubly occupancy of local Ni $3d_{x^{2}-y^{2}}$ orbital. As in cuprates, the holes' motion is strongly renormalized due to the onsite Coulomb
repulsion $U$.

Quite generally, the effective $t$-$J$-$K$ model can be replaced by the
one-band Hubbard model plus a hybridization term with additional conduction
electrons:
\begin{eqnarray}
H&=&\sum_{\mathbf{k}\sigma}E_{\mathbf{k}}d_{\mathbf{k}\sigma}^\dagger
d_{\mathbf{k}\sigma}+U\sum_i n^d_{i\uparrow}n^d_{i\downarrow}  \notag \\
&&+\sum_{\mathbf{k}\sigma}\epsilon_{\mathbf{k}}c_{\mathbf{k}\sigma}^\dagger
c_{\mathbf{k}\sigma}+\sum_{\mathbf{k}\sigma}V_{\mathbf{k}}\left(d_{\mathbf{k}%
\sigma}^\dagger c_{\mathbf{k}\sigma}+h.c.\right).
\end{eqnarray}
The model may also be viewed as a periodic Anderson model with dispersive $d$
bands. It allows for a better treatment of charge fluctuations of the Ni 3$%
d_{x^2-y^2}$ orbitals, in particular for large Sr doping or small Coulomb
interaction. In this work, we only consider the minimal $t$-$J$-$K$ model
and show that it can already capture some main physics of the nickelates.

\subsection{Global phase diagram}

As is in heavy fermion systems, the $t$-$J$-$K$ model contains two competing
energy scales that support different ground states. The Heisenberg
superexchange $J$ favors an antiferromagnetic long-range order, while the
Kondo coupling $K$ tends to screen the local spins and form a nonmagnetic
ground state. In nickelates, due to the large charge transfer energy between
O-$2p$ and Ni-$3d_{x^{2}-y^{2}}$ orbitals, $J$ is expected to be smaller
that that (about $100$ meV) in cuprates. Raman scattering measurements
confirmed this expectation and estimated $J\approx 25\,$meV in bulk NdNiO$_2$
\cite{Fu2019}. First-principles calculations also suggested $J$ of the order
of 10 meV \cite{Hanghui2020}. There is at present no direct measurement of
the Kondo interaction $K$. However, from the observed resistivity minimum at
70-100 K in NdNiO$_2$ and LaNiO$_2$ \cite{Li2019,Ikeda2016}, $K$ may be
roughly estimated to be of the order of a few hundred meV \cite{Yang2008,
zhang-yang-zhang, Hanghui2020}. Thus for undoped nickelates, the Kondo
coupling is a relatively large energy scale.

We may propose a global phase diagram starting from an antiferromagnetic
ground state. The self-doping introduces equal numbers of conduction
electrons and holes. The conduction electrons tend to form Kondo singlets
with local spins due to the large $K$. Both tend to suppress the long-range
AF order and causes a paramagnetic ground state. But because of the small
number of conduction electrons, local spins cannot be fully Kondo screened
to become delocalized. Thus, instead of a heavy fermion metal, we are
actually dealing with a low carrier density Kondo system at low
temperatures. Due to insufficient Kondo screening, the resistivity
exhibits insulating-like behavior (upturn) because of incoherent Kondo
scattering, which is typical for low carrier density Kondo systems and has
been observed previously in CeNi$_{2-\delta}$(As$_{1-x}$P$_{x}$)$_{2}$ \cite%
{Chen2019} and NaYbSe$_2$ \cite{Xu2021}.

\begin{figure}[t]
\centerline{{\includegraphics[width=.45\textwidth]{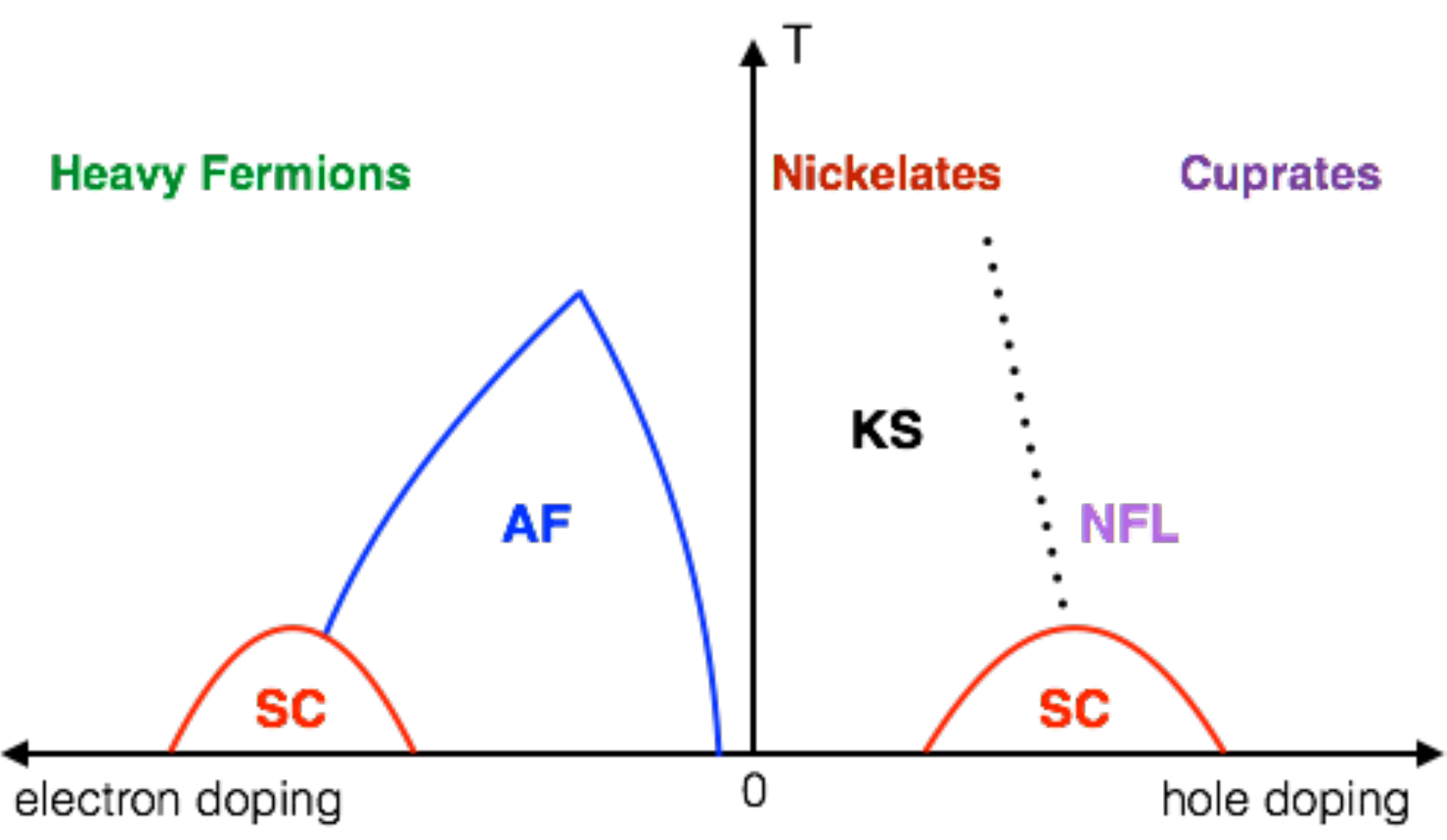}}}
\caption{A schematic phase diagram of the $t$-$J$-$K$ model with electron
and hole doping. KS stands for incoherent Kondo scattering or Kondo
semimetal. Whether or not the AF phase may exist depends on details of
electron and hole densities introduced by doping.}
\label{figGlobal}
\end{figure}

Upon Sr (hole) doping, the number of conduction electrons may be reduced,
while that of holes increases. The Kondo physics may be suppressed and
replaced by the usual $t$-$J$ model for large hole doping, resembling the
physics of cuprates. On the other hand, for electron doping, we may expect
to first recover the AF long-range order with reduced hole density, and then
with increasing conduction electrons and Kondo screening, the AF order will be 
suppressed again and the system turns into a heavy fermion metal with sufficient
electron doping. SC may emerge around the quantum critical point. Whether or not the AF phase may actually exist depends on
how doping changes the fraction of electron and hole carriers. But in bulk Nd%
$_{1-x}$Sr$_x$NiO$_2$, NMR experiment has revealed short-range glassy AF
ordering, supporting the possible existence of antiferromagnetism \cite%
{Cui2021}. Figure \ref{figGlobal} summarizes possible ground states of the
model on the temperature-doping plane, showing a connection between the
heavy fermion and cuprate physics on two ends and the nickelates in between.
However, it should be noted that current experiment on ``overdoped"
nickelate superconductors found a weak insulator rather than a Fermi liquid
as in heavily hole-doped cuprates \cite{Zeng2020}. How exactly holes are
doped in Nd$_{1-x}$Sr$_x$NiO$_2$ remains an open question.

\subsection{Low energy excitations}

As shown in the phase diagram, the paramagnetic region (KS) is
responsible for undoped or low doped nickelates. In this case, we have a
small number ($n_{c}$) of conduction electrons per Ni-site and $n_{c}+p$
empty nickel sites (holons) on the NiO$_{2}$ plane, where $p$ is the hole
doping ratio. In the large $K$ limit and at zero temperature, conduction
electrons form Kondo singlets or doublons with local Ni spins. We may then
derive an effective low-energy Hamiltonian in terms of doublons, holons, and
localized spins, to describe a doped Mott metallic state with Kondo
singlets. A cartoon picture is given in Fig.~\ref{figLattice}.

For this, we first introduce the pseudofermion representation for the spin-1/2 local moments: 
\begin{equation*}
S_{j}^{+}=f_{j\uparrow }^{\dagger }f_{j\downarrow },S_{j}^{-}=f_{j\downarrow
}^{\dagger }f_{j\uparrow },S_{j}^{z}=\frac{1}{2}\left( f_{j\uparrow
}^{\dagger }f_{j\uparrow }-f_{j\downarrow }^{\dagger }f_{j\downarrow
}\right),
\end{equation*}
where $f_{j\sigma }$ is a fermionic operator for the spinon on site $j$. The Ni $3d_{x^{2}-y^{2}}$ electron operator is given by $d_{j\sigma
}=h_{j}^{\dagger }f_{j\sigma }$ with a local constraint, $h_{j}^{\dagger
}h_{j}+\sum_{\sigma }f_{j\sigma }^{\dagger}f_{j\sigma }=1$, if we ignore
double occupancy. Here $h_{j}^{\dagger }$ is a bosonic operator creating a holon on site $j$.

The doublon operators for the on-site Kondo spin singlet and triplets may be
defined as
\begin{eqnarray*}
b_{j0}^{\dagger } &=&\frac{1}{\sqrt{2}}\left( f_{j\uparrow }^{\dagger
}c_{j\downarrow }^{\dagger }-f_{j\downarrow }^{\dagger }c_{j\uparrow
}^{\dagger }\right) ;\text{ } \\
b_{j1}^{\dagger } &=&f_{j\uparrow }^{\dagger }c_{j\uparrow }^{\dagger },\
b_{j2}^{\dagger }=\frac{1}{\sqrt{2}}\left( f_{j\uparrow }^{\dagger
}c_{j\downarrow }^{\dagger }+f_{j\downarrow }^{\dagger }c_{j\uparrow
}^{\dagger }\right) ,\ b_{j3}^{\dagger }=f_{j\downarrow }^{\dagger
}c_{j\downarrow }^{\dagger }.
\end{eqnarray*}
The Kondo term then becomes
\begin{equation}
\frac{K}{2}\sum_{j\alpha ;\sigma \sigma ^{\prime }}S_{j}^{\alpha }c_{j\sigma
}^{\dagger }\tau _{\sigma \sigma ^{\prime }}^{\alpha }c_{j\sigma ^{\prime }}=%
\frac{K}{4}\sum_{\mu =1}^{3}b_{j\mu }^{\dagger }b_{j\mu }-\frac{3K}{4}%
\sum_{j}b_{j0}^{\dagger }b_{j0},
\end{equation}%
which describes the doublon formation on each site, namely, the Kondo
singlet or triplet pair formed by one conduction electron with a localized
spinon. We see that the triplet pair costs a higher energy of $K$.
Similarly, there may also exist three-particle states with one localized
spinon and two conduction electrons on the same site, $e_{j\sigma }^{\dagger
}=f_{j\sigma }^{\dagger }c_{j\uparrow}^{\dagger }c_{j\downarrow }^{\dagger }$%
, or one-particle states with one unpaired spinon only, $\widetilde{f}%
_{j\sigma }=\left( 1-n_{j}^{c}\right) f_{j\sigma }$.

\begin{figure}[t]
\centerline{{\includegraphics[width=.45\textwidth]{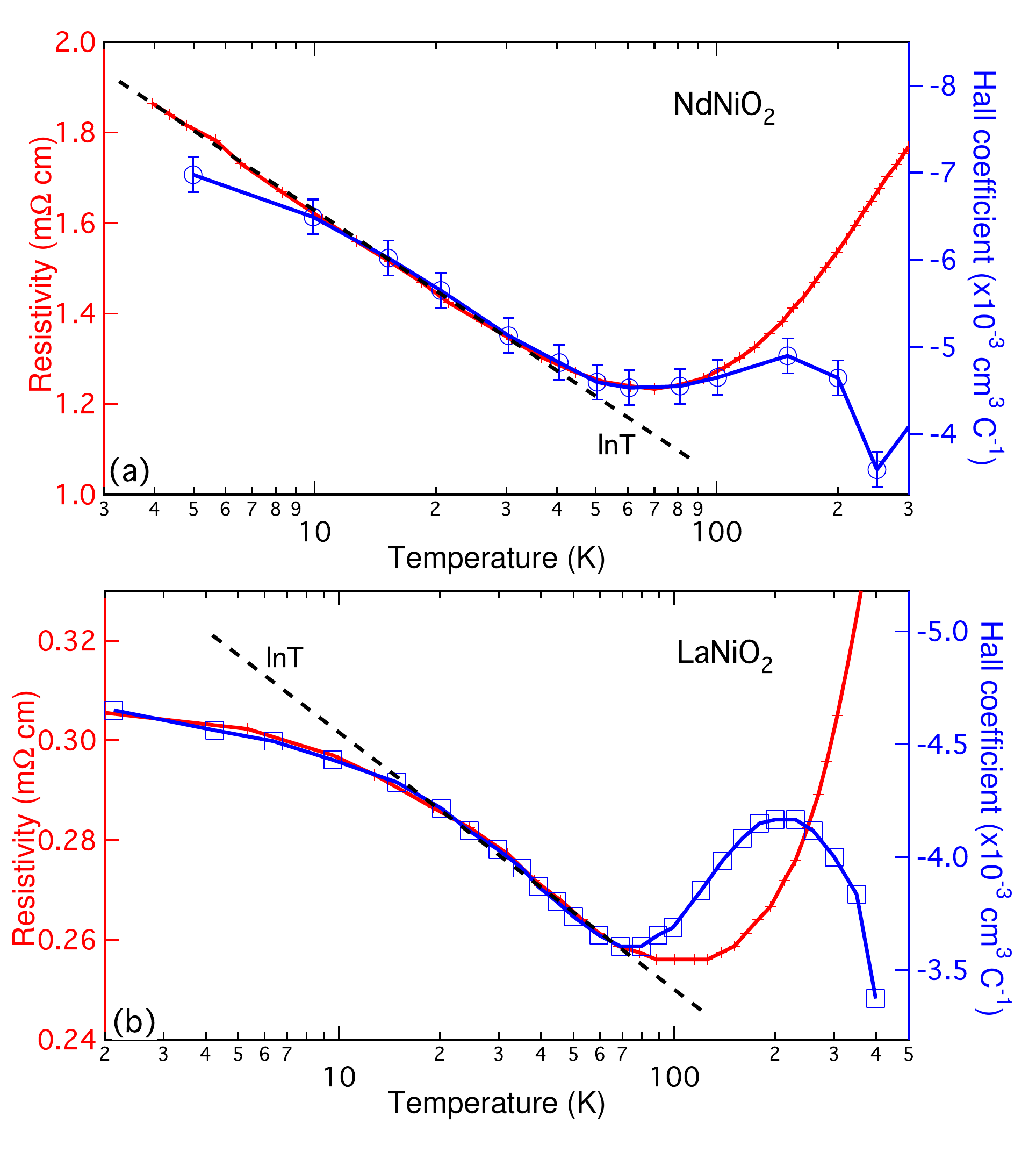}}}
\caption{Logarithmic temperature dependence of the resistivity (red color)
and the Hall coefficient (blue color) at low temperatures for (a) NdNiO$_{2}$
with the experimental data adopted from Ref.~\protect\cite{Li2019}; (b) LaNiO%
$_{2}$ reproduced from Ref.~\protect\cite{Ikeda2016}. The dashed lines are
the $\mathrm{ln} T$ fits. Figure adapted from Ref. \protect\cite%
{zhang-yang-zhang}. Copyright 2020 by the American Physical Society.}
\label{figRH}
\end{figure}

Following Refs. \cite{Lacroix1985,Sigrist1992}, we first rewrite the
Hamiltonian in terms of these new operators and then eliminate all
high-energy terms containing $b_{j\mu }$ ($\mu =1,2,3$) and $e_{j\sigma }$
using canonical transformation while keeping only the on-site doublon ($%
b_{j0}$) and unpaired spinons ($\widetilde{f}_{j\sigma }$). This yields an
effective low-energy model with a simple form
\begin{eqnarray}
H_{\mathrm{eff}} &=&-t\sum_{\langle ij\rangle ,\sigma }\left( h_{i}%
\widetilde{f}_{i\sigma }^{\dagger }\widetilde{f}_{j\sigma }h_{j}^{\dagger
}+h.c.\right) +J\sum_{\langle ij\rangle }\widetilde{S}_{i}\cdot \widetilde{S}%
_{j}  \notag \\
&&-\frac{t_c}{2}\sum_{\langle ij\rangle ,\sigma }\left( b_{i0}^{\dagger }%
\widetilde{f}_{i\sigma }\widetilde{f}_{j\sigma }^{\dagger
}b_{j0}+h.c.\right) ,
\end{eqnarray}%
where the spin operators are $\widetilde{S}_{j}^{\alpha }=\sum_{\sigma
\sigma ^{\prime }}\widetilde{f}_{j\sigma }^{\dagger }\frac{1}{2}\tau
_{\sigma \sigma ^{\prime }}^{\alpha }\widetilde{f}_{j\sigma ^{\prime }}$
with a local constraint $h_{j}^{\dagger }h_{j}+b_{j0}^{\dagger
}b_{j0}+\sum_{\sigma }\widetilde{f}_{j\sigma }^{\dagger }\widetilde{f}%
_{j\sigma }=1$. Here only the nearest-neighbor hopping parameters $t$ and $t_c$ are considered for simplicity. For large but finite $K$, apart from some complicated
interactions, an additional term should also be included
\begin{equation}
H_{b}=-\frac{3}{4}\left( K+\frac{t_c^{2}}{K}\right) \sum_{j}b_{j0}^{\dagger
}b_{j0}+\frac{5t_c^{2}}{12K}\sum_{\langle ij\rangle }b_{i0}^{\dagger
}b_{i0}b_{j0}^{\dagger }b_{j0},
\end{equation}%
which describes the doublon condensation.

This effective Hamiltonian describes the ground state of nickelates at zero
or low doping. It is similar to the usual $t$-$J$ model for cuprates \cite%
{ZhangRice}, but includes two types of mobile quasiparticles: doublons
(Kondo singlets) and holons. It is now clear why the self-doping can
efficiently suppress the AF long-range order to yield a paramagnetic ground
state in nickelates. At high temperatures, doublons become deconfined,
causing incoherent Kondo scattering in transport measurements. This is
confirmed by the resistivity replotted in Fig.~\ref{figRH} as a
function of temperature for both NdNiO$_{2}$ and LaNiO$_{2}$. Unlike cuprates, the resistivity exhibits metallic behavior at high temperature but
shows an upturn below about 70 K. If we put the data on a linear-log scale,
we find that the upturn follows exactly a logarithmic temperature ($\ln T$)
dependence over a large temperature range for both compounds, which is a
clear evidence for incoherent Kondo scattering typical for low carrier
density Kondo systems \cite{HewsonBook}. The saturation at very low
temperatures is an indication of Kondo screening. This Kondo scenario is also 
supported by the Hall measurement. In both compounds, the Hall coefficient $R_{H}$ exhibits non-monotonic temperature dependence. It approaches a negative constant at high temperatures due to the contribution of conduction electrons, but exhibits the same $\ln T$ dependence at low
temperatures. The linear relation $R_{H}\propto \rho $ is an indication of
skew scattering by localized magnetic impurities in typical Kondo systems
\cite{Fert1987,Nagaosa2010}. 

An alternative explanation for the resistivity upturn is weak localization, where disordered holes in the NiO$_{2}$ plane may also give rise to a logarithmic correction. However, this explanation is not supported by the corresponding correction to the Hall coefficient and magnetoresistance.

\subsection{Superconductivity}

Experimentally, superconductivity was first observed to emerge and have an
onset temperature of $14.9$ K in $20$\% Sr doped NdNiO$_{2}$ thin films
deposited on SrTiO$_3$ substrates \cite{Li2019}. From the $t$-$J$-$K$ model,
one may naively expect that sufficiently large doping may deplete conduction
electrons and increase the number of holons, driving the system to an
effective $t$-$J$ model resembling that in cuprates. As a result, $d$-wave
superconductivity may arise due to the superexchange interaction between Ni 3%
$d_{x^2-y^2}$ electrons. However, this is not the whole truth. With
increasing Sr doping, the ordinary Hall coefficient at high temperatures
becomes smaller in magnitude but remains negative even in Nd$_{0.8}$Sr$%
_{0.2} $NiO$_{2}$ \cite{Li2019}, which cannot be explained by a single
carrier model but rather indicates a cancellation of electron and hole
contributions. Thus, conduction electrons should still be present even at $%
20 $\% Sr doping. It is thus anticipated that superconductivity in
nickelates may be affected by the presence of conduction electrons and their
Kondo hybridization with Ni 3$d_{x^2-y^2}$ electrons.

\begin{figure}[tbp]
\centering
\includegraphics[width=\columnwidth]{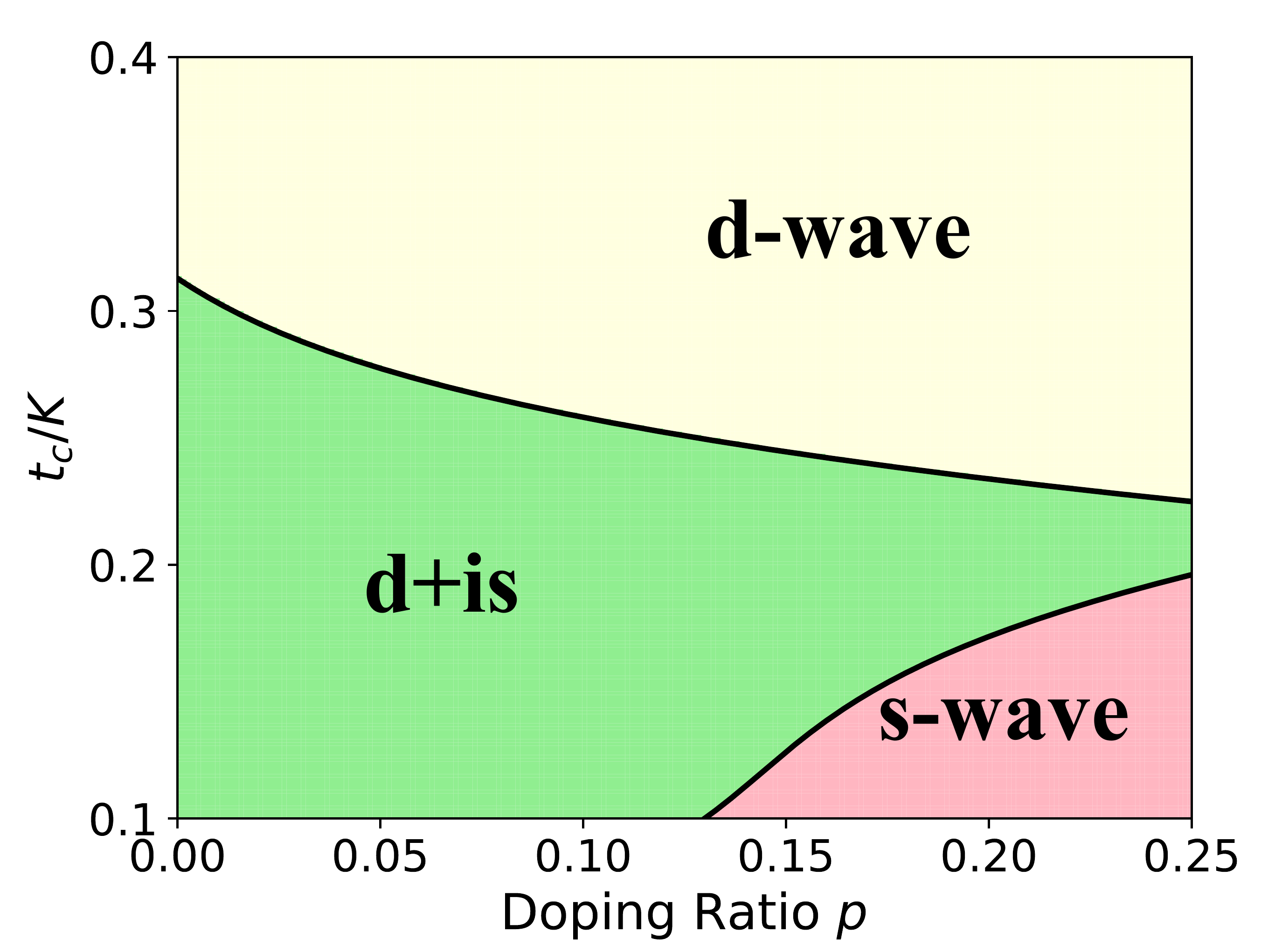}
\caption{Phase diagram of the superconductivity. At small hole concentration
$p$, the pairing symmetry is primarily ($d+is$)-wave SC. At large doping $p$%
, the pairing is either $s$-wave SC for small hopping $t_{c}/K$ or $d$-wave
SC for large hopping $t_{c}/K$. Figure adapted from Ref.
\protect\cite{Wang2020}. Copyright 2020 by the American Physical Society.}
\label{figSC}
\end{figure}

The pairing symmetry can be studied by using the renormalized
mean-field theory (RMFT) \cite{RMFT,Wang2020}, which had successfully predicted the $d
$-wave superconductivity in the $t$-$J$ model for cuprates and can well
describe the Fermi liquid similar to the slave-boson mean field theory. Our
numerical calculations of superconductivity on the generalized $t$-$J$-$K$
model have yielded a typical superconducting phase diagram in Fig.~\ref%
{figSC}. The pairing symmetry is found to depend on the hole concentration $p
$ and the effective strength of the Kondo hybridization controlled by the
conduction electron hopping ($t_{c}/K$) \cite{zhang-yang-zhang}. For
simplicity, we have taken the Kondo coupling $K$ as the energy unit ($K=1$),
and set the AF Heisenberg spin exchange $J=0.1$. Both the nearest-neighbor
hopping $t=0.2$ and the next-nearest-neighbor hopping $t^{\prime }=-0.05$ are taken into consideration  
for Ni 3$d$ electrons. The density of conduction electrons is taken to be $%
n_{c}=0.1$, while their nearest-neighbor hopping $t_{c}$ is chosen as a
tuning parameter.

For all doping, we find a dominant $d$-wave pairing symmetry for large $%
t_c/K $ or small $K$. But for low doping and moderate $t_c/K$, we obtain a $%
\left( d+is\right) $-wave pairing state that breaks the time-reversal
symmetry. This is different from the conventional picture based purely on the
$t$-$J$ model for cuprates, where the Heisenberg superexchange interaction
favors $d$-wave pairing \cite{Thomale}. The $s$-wave pairing should be
ascribed to the Kondo coupling, which is an on-site spin exchange between
localized and conduction electrons \cite{Bodensiek2013}. This is supported
by the extended $s$-wave solution at large doping for small $t_c/K$ or
strong Kondo coupling. It is the combination of both effects that gives rise
to the special $\left( d+is\right) $-wave superconductivity and represents a
genuine feature of the nickelate superconductivity differing from cuprates
or heavy fermions.

Details of RMFT calculations are explained as follows \cite{Wang2020}. We first introduce 
three Gutzwiller renormalization factor to approximate the operator that
projects out the doubly occupied states: $g_{t}=n_{h}/(1+n_{h})$ for the
hopping $t$ and $t^{\prime }$, $g_{J}=4/(1+n_{h})^{2}$ for the superexchange
$J$, and $g_{K}=2/(1+n_{h})$ for the Kondo coupling $K$. We then define four
mean-field order parameters to decouple the Heisenberg superexchange and
Kondo terms:
\begin{eqnarray*}
\chi _{ij} &=&\langle d_{i\uparrow }^{\dagger }d_{j_{\uparrow
}}+d_{i\downarrow }^{\dagger }d_{j\downarrow }\rangle ,\quad B=\frac{1}{%
\sqrt{2}}\langle d_{j\uparrow }^{\dagger }c_{j\downarrow }^{\dagger
}-d_{j\downarrow }^{\dagger }c_{j\uparrow }^{\dagger }\rangle , \\
\Delta _{ij} &=&\langle d_{i\uparrow }^{\dagger }d_{j_{\downarrow
}}^{\dagger }-d_{i\downarrow }^{\dagger }d_{j\uparrow }^{\dagger }\rangle
,\quad D=\frac{1}{\sqrt{2}}\langle c_{j\uparrow }^{\dagger }d_{j\uparrow
}+c_{j\downarrow }^{\dagger }d_{j\downarrow }\rangle .
\end{eqnarray*}%
The resulting mean-field Hamiltonian has a simple bilinear form in the
momentum space,
\begin{equation}
\mathcal{H}_{\text{mf}}=\sum_{\mathbf{k}}\Psi _{\mathbf{k}}^{\dag }\left(
\begin{array}{cccc}
\chi (\mathbf{k}) & K_{D} & \Delta ^{\ast }(\mathbf{k}) & K_{B}^{\ast } \\
K_{D}^{\ast } & \epsilon(\mathbf{k}) & K_{B}^{\ast } & 0 \\
\Delta (-\mathbf{k}) & K_{B} & -\chi (-\mathbf{k}) & -K_{D}^{\ast } \\
K_{B} & 0 & -K_{D} & -\epsilon(-\mathbf{k})%
\end{array}%
\right) \Psi _{\mathbf{k}},
\end{equation}%
where we have introduced the Nambu spinors $\Psi _{\mathbf{k}}^{\dagger
}=(d_{\mathbf{k}\uparrow }^{\dagger },c_{\mathbf{k}\uparrow }^{\dagger },d_{-%
\mathbf{k\downarrow }},c_{-\mathbf{k}\downarrow })$ and defined the matrix
elements
\begin{eqnarray}
\chi (\mathbf{k}) &=&-\sum_{\alpha }\left( tg_{t}+\frac{3}{8}Jg_{J}\chi
_{\alpha }\right) \cos (\mathbf{k}\cdot \alpha )  \notag \\
&&-t^{\prime }g_{t}\sum_{\delta }\cos (\mathbf{k}\cdot \delta )+\mu _{1},
\notag \\
\epsilon(\mathbf{k}) &=&-t_{c}\sum_{\alpha }\cos (\mathbf{k}\cdot
\alpha )+\mu _{2},  \notag \\
\Delta (\mathbf{k}) &=&-\frac{3}{8}Jg_{J}\sum_{\alpha }\Delta _{\alpha }\cos
(\mathbf{k}\cdot \alpha ),  \notag \\
K_{D} &=&-\frac{3}{4}g_{K}K\frac{D}{\sqrt{2}},K_{B}=-\frac{3}{4}g_{K}K\frac{B%
}{\sqrt{2}}.
\end{eqnarray}%
Here $\alpha $ denotes the vectors of the nearest-neighbor lattice sites and
$\delta $ stands for those of the next-nearest-neighbor sites. $\mu _{1}$
and $\mu _{2}$ are chemical potentials fixing the numbers of the constrained
electrons $d_{i\sigma }$ and conduction electrons $c_{i\sigma }$,
respectively.

\begin{figure}[tbp]
\centering
\includegraphics[width=\columnwidth]{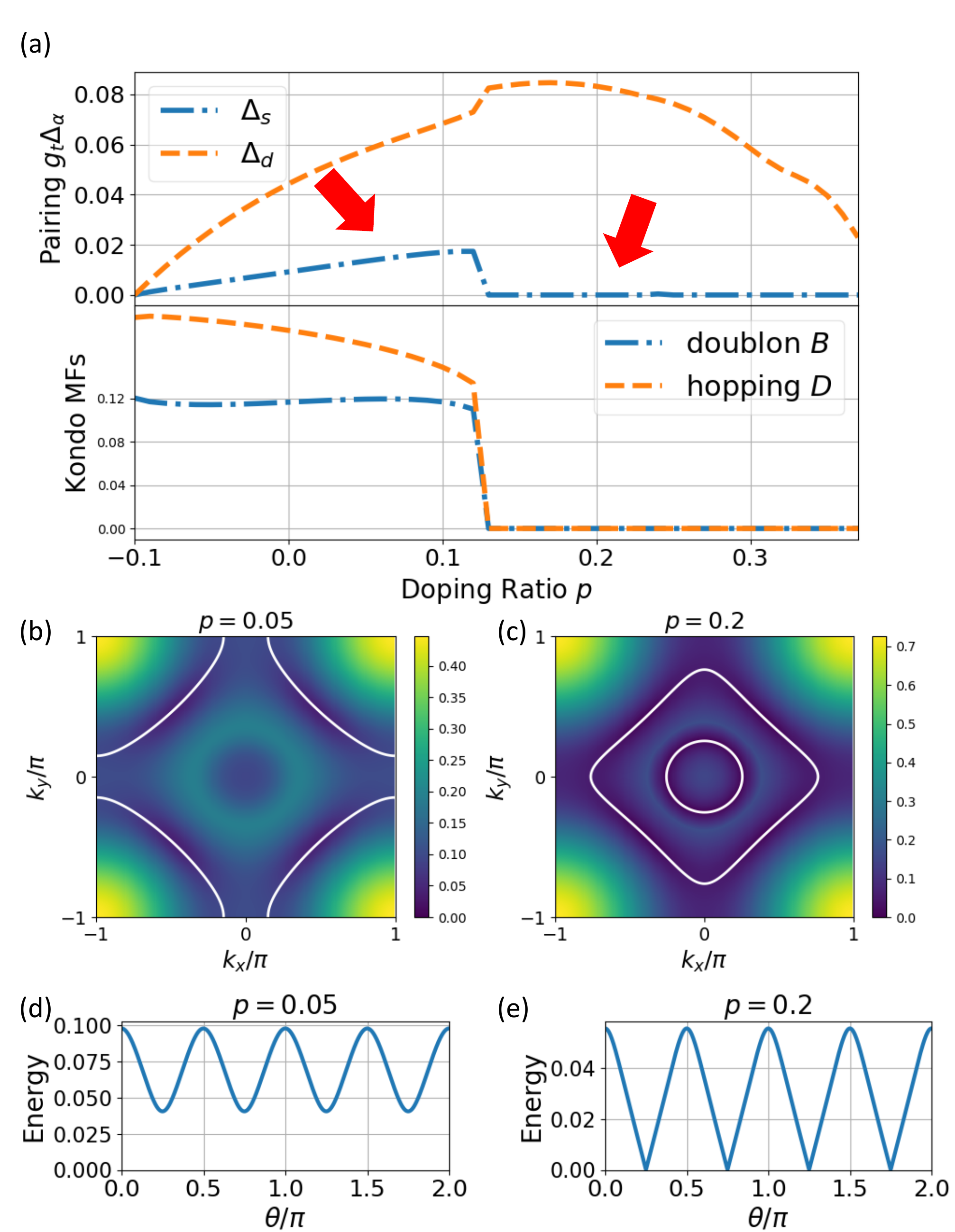}
\caption{RMFT results for $t_{c}/K=0.25$. (a) The mean-field parameters as a
function of doping for $g_{t}\Delta$ (upper panel) and $B$ and $D$ (lower
panel). (b) and (c) show the quasiparticle excitation energy (background)
and the Fermi surface (white solid line) defined as the minimal excitation
energy at $p=0.05$ ($d+is$)-wave and $p=0.2$ ($d$-wave) as marked by the
arrows in (a). (d) and (e) displayed the respective superconducting gap
along the Fermi surface. Figure adapted from Ref. \protect\cite%
{Wang2020}. Copyright 2020 by the American Physical Society.}
\label{figRMFT}
\end{figure}

The above mean-field Hamiltonian can be diagonalized using the Bogoliubov
transformation, $\left( d_{\mathbf{k}\uparrow },c_{\mathbf{k}\uparrow },d_{-%
\mathbf{k}\downarrow }^{\dagger },c_{-\mathbf{k}\downarrow }^{\dagger
}\right) ^{T}=U_{\mathbf{k}}\left( \alpha _{\mathbf{k}\uparrow },\beta _{%
\mathbf{k}\uparrow },\alpha _{-\mathbf{k}\downarrow }^{\dagger },\beta _{-%
\mathbf{k}\downarrow }^{\dagger }\right) ^{T}$. The ground state is given by
the vacuum of the Bogoliubov quasiparticles $\{\alpha _{\mathbf{k}\sigma
}^{\dagger },\beta _{\mathbf{k}\sigma }^{\dagger }\}$, which in turn yields
the self-consistent equations for the mean-field order parameters. We will
not go into more details here, but only mention that the mean-field
self-consistent equations can be solved numerically and yield two dominant
pairing channels of $s$ and $d$-waves as shown in the phase diagram Fig.~\ref%
{figSC}. Typical results for the mean-field parameters are plotted in Fig. %
\ref{figRMFT}(a) as a function of the doping ratio $p$ for $t_{c}/K=0.25$.
For clarity, we have defined $\Delta _{s}=|\Delta_{x}+\Delta _{y}|/2$ and $%
\Delta _{d}=|\Delta _{x}-\Delta _{y}|/2$ to represent the respective pairing
amplitudes of $s$ and $d$ channels. We see a clear transition from the mixed
$\left( d+is\right) $-wave SC to the pure $d$-wave SC.

Antiferromagnetic spin fluctuations have been observed in bulk Nd$_{1-x}$Sr$%
_{x}$NiO$_{2}$ by NMR \cite{Cui2021} and may also exist and play the role of
pairing glues in thin films. Hence the presence of a dominant $d$-wave
pairing at large doping is expected from the experience in cuprates.
However, our results also suggest several additional features of the
nickelate superconductivity that are not present in cuprates and may be
examined in experiment. First, for sufficiently large Kondo coupling $K$,
the $(d+is)$-wave SC in the low doping region breaks the time reversal
symmetry and as shown in Fig. \ref{figRMFT}(d), has a nodeless gap which is
distinctly different from the usual $d$-wave pairing with nodes along the
diagonal direction. Second, we predict a quantum phase transition between
this gapped $(d+is)$-wave SC to the nodal $d$-wave superconductivity with
increasing hole doping. These features can be detected by scanning
tunneling, penetration depth, or $\mu $SR experiment and serve as a support
for our theory. We remark that conduction electrons play an important role
in our theory of nickelate superconductivity, which is not possible in the
single-band Mott picture. The importance of electron pockets is in fact
supported by experimental measurements of the upper critical field \cite%
{Xiang2021,Wang2021}.

Recent single particle tunneling experiment \cite{HaihuWen} on
superconducting nickelate thin films have also observed two distinct types
of spectra: a V-shape feature with a gap maximum of 3.9 meV, a U-shape
feature with a gap of about 2.35 meV, and some spectra with mixed
contributions of these two components. If we attribute their different
observations to different hole concentrations due to possible surface
effect, the two types of spectra may correspond exactly to the two pairing
states in our theory. In this sense, the scanning tunneling spectra have
provided a supportive evidence for our theoretical prediction of multiple
superconducting phases. Of course, the $(d+is)$-wave pairing might not exist
in real materials if the Kondo coupling $K$ is too weak or $t_{c}/K$ is too
large. In that case, as is seen in Fig. \ref{figSC}, $d$-wave pairing may
become dominant on the hole Fermi surface, but electron pockets may still
have nodeless gap depending on their position in the Brillouin zone.

\subsection{Quantum criticality}

As shown in Fig.~\ref{figRMFT}, for moderate $t_c/K$, the SC transition from
$(d+is)$ to $d$-wave is accompanied with vanishing Kondo mean-field
parameters $B$ and $D$, which implies a breakdown of the Kondo hybridization
in the large doping side. Correspondingly, the Fermi surface structures also
change from a large hole-like Fermi surface around four Brillouin zone
corners at low doping to two separate electron-like Fermi surfaces (from
decoupled charge carriers) around the Brillouin zone center at large doping.
These results may be compared with the Hall experiments in Nd$_{1-x} $Sr$%
_{x} $NiO$_{2}$ \cite{Li2020,Zeng2020}, which revealed a crossover line of
sign change in the temperature-doping phase diagram near the maximal $T_c$.
The line marks a potential change in the Fermi surfaces and resembles that
observed in some heavy fermion systems owing to the delocalization of
localized moments \cite{Yang2017}. It is thus attempted to link the
experiment with our theoretical proposals and predict a zero-temperature
quantum critical point with the SC transition and the Fermi surface change
near the crossover line, although it should be cautious that they take place
in different temperature region.

\begin{figure}[t]
\centerline{{\includegraphics[width=.5\textwidth]{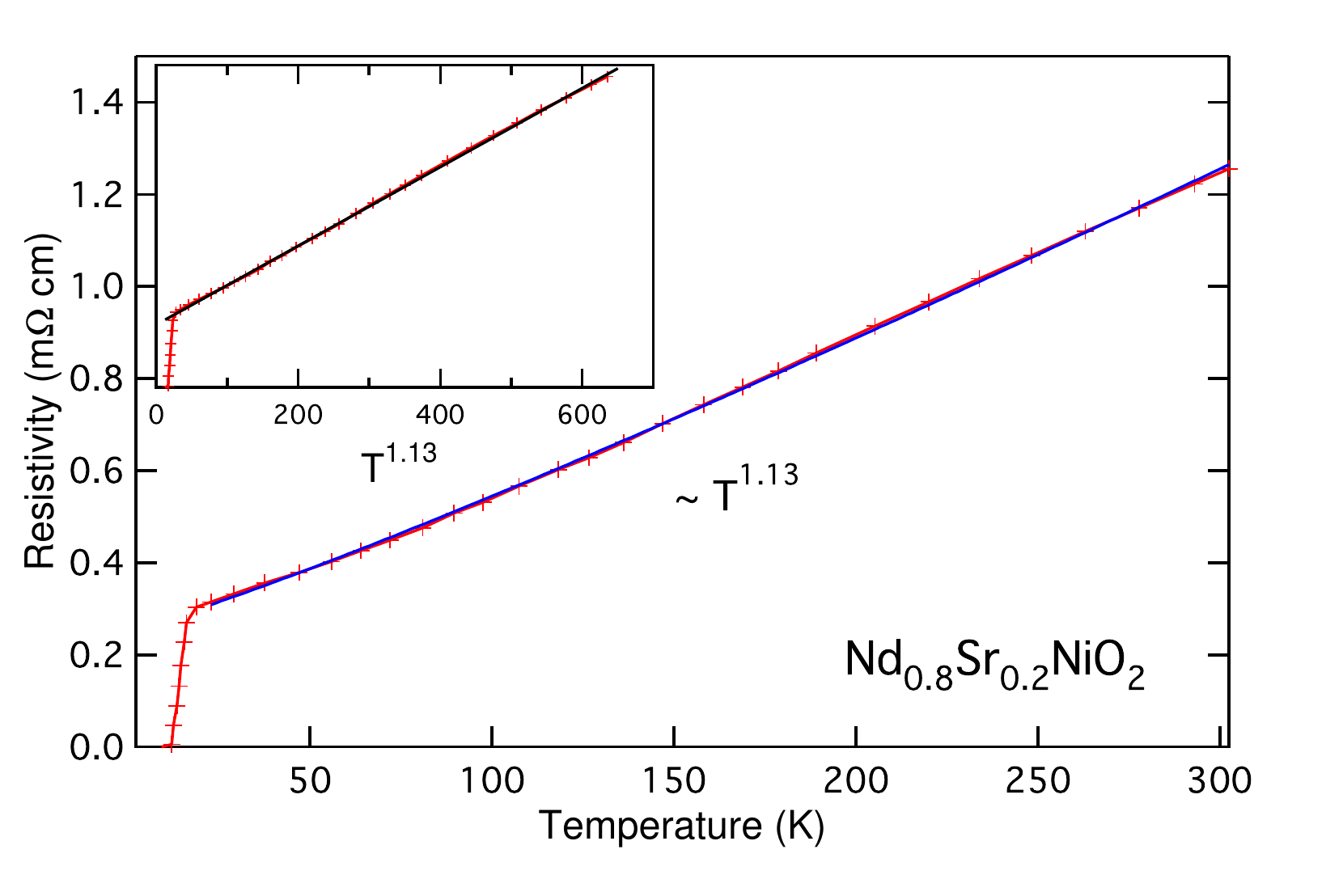}}}
\caption{Power-law temperature dependence of the electric resistivity above $%
T_c=15\,$K up to the room temperature for Nd$_{0.8}$Sr$_{0.2}$NiO$_{2}$. The
experimental data were reproduced from Ref.~\protect\cite{Li2019}.}
\label{figQC}
\end{figure}

As a matter of fact, experiment has indeed observed quantum critical
behavior in the normal state above $T_c$. A tentative fit of the resistivity
in superconducting nickelate thin films has yielded power-law scaling with
temperature, namely $\rho \sim T^{\alpha }$, with $\alpha =1.1-1.3$ over a
wide range \cite{zhang-yang-zhang}. Figure \ref{figQC} gives an example of
the fit in Nd$_{0.8}$Sr$_{0.2}$NiO$_{2}$ and we obtain $\alpha\approx1.13$
from slightly above $T_c$ up to the room temperature. This reminds us the
strange metal above $T_c$ in optimal-doped cuprates and the non-Fermi liquid
in heavy fermion systems. Better numerical calculations are required in
order to establish the exact mechanism of this scaling.

\section{Future perspective}

We have introduced the picture of self-doped Mott insulator to describe the
recently discovered nickelate superconductors. The self-doping effect has
been generally accepted by the community and distinguishes nickelates from
cuprates. We further propose a Mott-Kondo scenario and an extended $t$-$J$-$K
$ model based on transport measurements and electronic structure
calculations. Our model bridges the usual Kondo lattice model for heavy
fermions and the $t$-$J$ model for cuprates, but shows unique features that
can only be understood as an interplay of both physics. Our theory provides
a natural explanation of the resistivity upturn in undoped nickelates at low
temperatures, and our calculations based on the $t$-$J$-$K$ model predict an
exotic $(d+is)$-wave superconductivity that breaks the time reversal symmetry
and a possible transition of the pairing symmetry at critical doping, around
which the normal state exhibits non-Fermi liquid behavior above $T_{c}$.
This implies that nickelate superconductors are a novel class of
unconventional superconductors. Thus, exploration of new physics based on
our theory will be an interesting direction for future investigations.

Currently there still exist different opinions on the effect of Sr doping.
Some argued that holes may occupy other Ni 3$d$ orbitals and the Hund
coupling may favor a high spin state ($S=1$). This scenario seems
inconsistent with joint analyses of XAS and RIXS experiments \cite{Rossi2020} and a number of other calculations \cite{Jiang2020}. Nevertheless,
our model allows for a straightforward multi-orbital extension by
considering a two-band Hubbard model of Ni $3d$ orbitals (or its projection
at large $U$) plus hybridization with additional conduction bands. So far,
bulk nickelates have not been found superconductive and many issues remain
to be answered both in theory and in experiment \cite{Gu2021}. Our proposal
of the self-doping effect, the Mott-Kondo scenario, and the $t$-$J$-$K$
model provides a promising starting basis for future investigations.

\section{Acknowledgments}

This work was supported by the National Natural Science Foundation of China
(11774401, 11974397, 12174429), the National Key Research and Development
Program of MOST of China (2017YFA0303103, 2017YFA0302902), and the Strategic Priority
Research Program of CAS (Grand No. XDB33010100).

\end{document}